\documentclass[fleqn,twoside]{article}
\usepackage{espcrc2}
\usepackage{graphicx}

\newcommand{\AmS}{{\protect\the\textfont2
  A\kern-.1667em\lower.5ex\hbox{M}\kern-.125emS}}

\title{$I=0$ scalar channel}

\author{
SCALAR Collaboration:
S.~Muroya\address
{Tokuyama Women's College, Tokuyama 745-8511, Japan},
A.~Nakamura\address[IMC]{IMC, Hiroshima University, Higashi-Hiroshima 739-8521, Japan}, 
C.~Nonaka\addressmark[IMC], 
M.~Sekiguchi\address{Faculty of Engineering, Kokushikan University, 
Tokyo 154-8515, Japan},
and 
H.~Wada\address{Laboratory of Physics, Nihon University, Chiba 274-8501, Japan}
}

\begin{document}

\begin{abstract}
Using lattice QCD with dynamical Wilson fermions,
we study $I=0$ and $J^{P}=0^{+}$ channel which is constructed
by  $\frac{1}{\sqrt{2}}(\bar{u}u+\bar{d}d)|0\rangle$, in order to search
for the $\sigma$ meson.
Our preliminary result shows that the connected and disconnected 
diagrams contribute to the $\sigma$ meson propagator in the same order.
\end{abstract}

\maketitle

\section{INTRODUCTION}

In QCD, the chiral symmetry is spontaneously broken (restored)
at the confinement (deconfinement) phase where sigma meson
plays an essential role. We do not doubt this mechanism, and yet 
no one is sure whether the sigma meson exists or not. 
It might have large mass and/or wide width,
or might be simply $\pi-\pi$ correlation effect. 
The light $\sigma$ meson had disappeared from the tables of 
Particle Data Group (PDG) for over 20 years.
However, the $I=0$ and $J^{PC}=0^{++}$ meson, 
``$f_0$(400-1200) or $\sigma$'', appeared bellow $1$ GeV mass 
region in PDG recently\cite{PDG}. 
This is probably because of recent $\pi$-$\pi$ scattering phase shift 
reanalyses; Especially, Igi and Hikasa constructed a general 
model-independent framework, 
which respects the  analyticity, unitarity and crossing symmetry
together with chiral symmetry low energy theorem,
to describe the $\pi\pi$ elastic
scattering below $1$ GeV mass region and investigated the existence of 
$\sigma$ meson \cite{Igi}. 
See Ref.\cite{Kunihiro} for a good review on the situation and
physical meaning of the sigma meson.

Now it is very desirable to investigate whether  $\sigma$ meson 
appears as a pole based on lattice QCD.
Using the quenched approximation, Alfold and Jaffe discussed the possibility 
of the light scalar mesons as $\bar{q}^2q^2$ states rather than $\bar{q}q$ 
\cite{Alfold}.
McNeile and Michael computed the 
mixed iso-singlet scalar masses of $q\bar{q}$ and glueball states
in two kind of situation, i.e,  with and without the dynamical quark 
effects \cite{McNeile}. The $\sigma$ meson masses which 
are obtained with consideration of the dynamical 
quark effects are much lower than the quenched results. 
The goal of our project is to conclude whether the $\sigma$ 
meson exists or not below $1$ GeV in QCD.

\section{$\sigma$ PROPAGATOR}

We construct $I=0$ scalar channel by $\sigma|0\rangle$. The
operator $\sigma$ is given as
\begin{equation}
\sigma(x) \equiv
 \sum_{c=1}^3\sum_{\alpha=1}^4 
\frac{\bar{u}_\alpha^c(x)u_\alpha^c(x)+\bar{d}_\alpha^c(x)d_\alpha^c(x)}
{\sqrt{2}},
\end{equation}
where $u$ and $d$ are the $u$-quark and $d$-quark Dirac spinors, 
respectively. 
This operator has the same quantum number as the vacuum, 
$I=0$ and $J^P=0^+$. 
The indices $c$ and $\alpha$ denote color and Dirac spinor indices, 
respectively. The $\sigma$ meson propagator is given by,
\begin{eqnarray}
 G(y,x)  =  - \langle {\mbox Tr} W^{-1}(x,y) W^{-1}(y,x) \rangle \nonumber \\
     + 2 \langle {\mbox Tr} W^{-1}(y,y) {\mbox Tr} W^{-1}(x,x) 
\rangle \nonumber \\
     - 2 \langle {\mbox Tr} W^{-1}(y,y) \rangle \langle {\mbox Tr} 
W^{-1}(x,x) \rangle .
\label{Eq-propa}
\end{eqnarray} 
\noindent
Here "Tr" represents summation over color and Dirac spinor indices
and
$W^{-1}$ is $u$, $d$ quark propagator.
The third term of Eq.(\ref{Eq-propa}), $\langle\sigma(y)\rangle\langle\sigma(x)\rangle$, 
corresponds to the subtraction of vacuum contribution.\footnote{
$\langle\sigma\rangle$ is nothing but $\langle\bar{\psi}\psi\rangle$.} 
The second and the third terms are the same order and 
the high precision numerical simulations and careful analyses are 
required.

\section{NUMERICAL SIMULATIONS}

We calculate the $\sigma$ propagator by using 
Hybrid Monte Carlo algorithm.
We use $Z_2$ noise method to calculate the disconnected diagrams. 
The $1000$ random $Z_2$ numbers are generated.
The two-flavors Wilson fermion is simulated on the $8^3\times 16$
lattice. 
Based on ref.\cite{CPPACS}, we set $\beta = 4.8$, $\kappa=0.1846$ 
($a=0.197(2)$ fm, $\kappa_c=0.19286(14)$) \cite{CPPACS}. 
After the thermalization trajectories,
$\sigma$ propagators are calculated on a configuration in every 
10 trajectories.

\subsection{NUMERICAL ACCURACY}

Since there are big numerical cancellation in 
$\langle\sigma(y)\sigma(x)\rangle -  
\langle\sigma(y)\rangle\langle\sigma(x)\rangle$,
we must be careful to controle the numerical accuracy.

\begin{figure}
\includegraphics[width=1.0\linewidth]{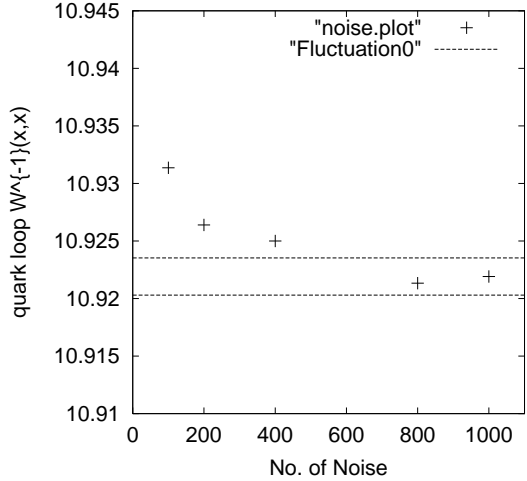}
\caption{ ${\mbox Tr} W^{-1}(x,x)$ evaluated by the noise
method as a function of the number of Z2 noise sources.}
\label{Fig-noise}
\end{figure}

In Fig.\ref{Fig-noise}, we show the values of $\sigma(x)$
for a typical configuration as a function of the number
of Z2 noise sources.  Dotted lines represent the required
accuracy to evaluate $\langle\sigma(y)\sigma(x)\rangle -
\langle\sigma(y)\rangle\langle\sigma(x)\rangle$.
From the figure, we may conclude that 1000 noise sources are
very safe and we do not suffer from any systematic error
due to the Z2 noise.

We have checked also the effect of CG solver tollerance, and
${\mbox Tr} W^{-1}(x,x)$ is very stable for $\epsilon < 10^{-10}$. 

\subsection{PROPAGATORS}
\begin{figure}
\includegraphics[width=1.0\linewidth]{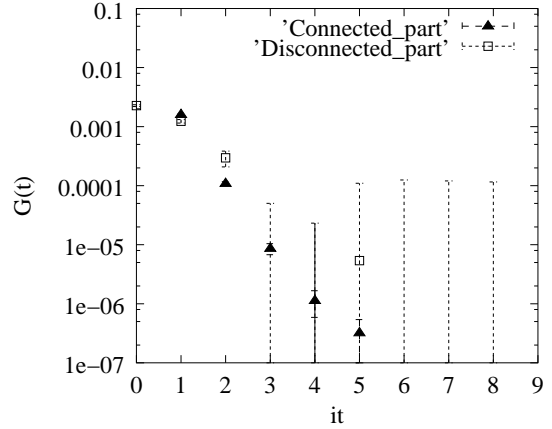}
\caption{The connected and disconnected propagator parts of $\sigma$ meson.}
\label{Fig-dc}
\end{figure}
\begin{figure}
\includegraphics[width=1.0\linewidth]{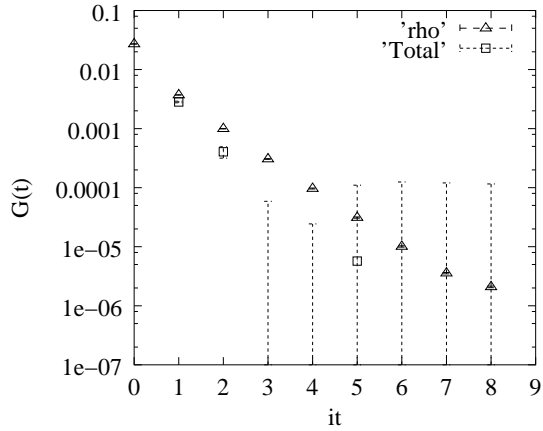}
\caption{The comparison between the $\sigma$ propagator and $\rho$ propagator.}
\label{Fig-sig_rs}
\end{figure}

As preliminary results, Fig.\ref{Fig-dc} shows the 
contribution of the connected and disconnected diagrams 
of the $\sigma$ propagator in the time direction. 
The connected diagram shows clear 
exponential damping.  
On the other hand, the disconnected diagram has large 
error bars.
However there exists the contribution of 
disconnected diagram to $\sigma$ meson propagator and its order is the same 
as connected diagram. 
Furthermore it can be possible that 
the existence of the disconnected diagram makes  
the $\sigma$ meson mass lighter.
This means that the estimation of the disconnected diagram 
is important to determine the property of $\sigma$ meson.
In order to argue $\sigma$ meson in detail, the improvement of the 
evaluation of the disconnected diagram and more statistics 
are indispensable.

In Fig.\ref{Fig-sig_rs} we compare the $\sigma$ meson propagetor with 
$\rho$ propagator. 
We can see that the $\sigma$ meson  mass could be the same order of $\rho$ 
meson mass, i.e., 
we obtain the suggestion of the existence of light $\sigma$ meson, 
though the error bar of $\sigma$ meson  propagator 
is very large. 

\section{CONCLUDING REMARKS}

We investigated the property of 
$I=0$ and $J^{PC}=0^{++}$ scalar meson ($\sigma$ meson) whose operator 
is $\frac{1}{\sqrt{2}}(\bar{u}u+\bar{d}d)$. 
In the $\sigma$ meson propagator the contribution of 
disconnected diagram is the same order of connected 
diagram; Quenched approximation is not reliable for
the investigation of the $\sigma$ meson. 
The evaluation of the disconnected diagram is done by 
using $Z_2$ noise method. 
A statistical error of $\sigma$ propagator which comes from the 
disconnected diagram mainly is large in the present stage. 
As preliminary results, we obtain the following properties of 
$\sigma$ propagator:
(1) Both the connected and disconnected parts equally
contribute to the $\sigma$ propagator.
(2) $\sigma$ meson could have mass of the same order of 
      the $\rho$ meson. 

It is necessary to generate much more gauge configurations 
and improve the statistical precision of 
the estimation of $\sigma$ propagator. 
We also plan to improve the source of the sigma propagators.

Furthermore we must investigate the mixing state of 
the $\sigma$ meson and glueball if we obtain the result that $\sigma$ meson 
mass is greater than $1$ GeV region\cite{Lee,Kisslinger}.

\section*{ACKNOWLEDGMENT}
We would like to  thank T. Kunihiro for the useful
discussions and encouragement. 
This work is supported by Grant-in-Aide for Scientific Research by
Monbu-Kagaku-sho, Japan (No.11440080, No. 12554008 and No. 13135216).
This work is performed by SX5 at RCNP, Osaka Univ.
One of the authors(C.N.) would like to
acknowledge the financial support by the Soryushi Shogakukai.

\end{document}